\documentclass{article}

\usepackage[a4paper, total={6in, 8in}]{geometry}
\usepackage[utf8]{inputenc}
\usepackage{amsmath}
\usepackage{amssymb}
\usepackage{makecell}
\usepackage{lscape}
\usepackage{graphicx}
\usepackage{changepage}
\usepackage{titlesec}
\usepackage{color, colortbl}
\usepackage{hyperref}
\newcommand{\footremember}[2]{%
	\footnote{#2}
	\newcounter{#1}
	\setcounter{#1}{\value{footnote}}%
}
\newcommand{\footrecall}[1]{%
	\footnotemark[\value{#1}]%
}
\DeclareMathOperator\erf{erf}
\definecolor{Gray}{gray}{0.9}
\newtheorem{theorem}{Theorem}
\titleformat{\paragraph}
{\normalfont\normalsize\bfseries}{\theparagraph}{1em}{}
\titlespacing*{\paragraph}
{0pt}{1.25ex plus 1ex minus .2ex}{1.5ex plus .2ex}

\title{Improving radiation dose estimation using the $\gamma$-H2AX biomarker}
\author{Dorota Młynarczyk\footremember{1}{Departament de Matemàtiques, Universitat Autònoma de Barcelona, Bellaterra, 08193, Barcelona.} \and Pedro Puig\footrecall{1}
\footnote{Centre de Recerca Matemàtica, Bellaterra, 08193 Barcelona.}\and Carmen Armero\footremember{2}{Departament d’Estadística i Investigació Operativa, Universitat de València, 46100 València.}
\and Virgilio Gómez-Rubio\footremember{3}{Department of Mathematics, School of Industrial Engineering, Universidad de Castilla-La Mancha, 02071 Albacete.}
\and Joan F. Barquinero\footremember{4}{Unitat d’Antropologia Biològica, Departament de Biologia Animal, Biologia Vegetal i Ecologia, Universitat Autònoma de Barcelona,
Bellaterra, 08193 Barcelona.} 
\and Mònica Pujol-Canadell\footrecall{4}
}

\date{}
\begin{document}
\maketitle

\begin{abstract}
To predict the health effects of accidental or therapeutic radiation exposure, one must estimate the radiation dose that person received. A well-known ionising radiation biomarker, phosphorylated $\gamma$-H2AX protein, is used to evaluate cell damage and is thus suitable for the dose estimation process. In this paper, we present new Bayesian methods that, in contrast to approaches where estimation is carried out at predetermined post-irradiation times, allow for uncertainty regarding the time since radiation exposure and, as a result, produce more precise results. We also use the Laplace approximation method, which drastically cuts down on the time needed to get results. Real data are used to illustrate the methods, and analyses indicate that the models might be a practical choice for the $\gamma$-H2AX biomarker dose estimation process.
\end{abstract}
\section{Introduction and motivation}
Ionising radiation is currently used for a variety of purposes, such as industrial radiography, energy production, and health diagnosis and treatment. This results in an increased risk of radiological accidents. When a radiological accident occurs, it is important to assess the dose of radiation absorbed by people affected to help decision-making measures and whether it is necessary to help with medical care. Biodosimetry refers to the use of several biological markers to quantify the exposure to radiation of any person suspected of being exposed. The present study focuses on the $\gamma$-H2AX protein, which is an accepted biomarker of dose exposure to ionising radiation \cite{rothkamm2009gamma}.

\begin{figure} 
\centering
\includegraphics[height=5.5cm]{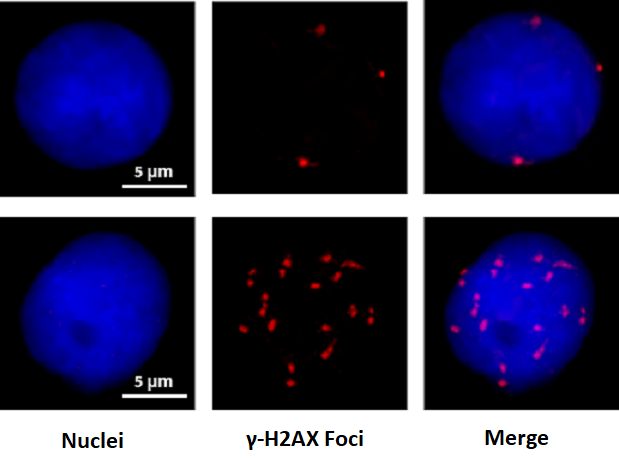}
\caption{Images of nuclei from peripheral blood mononucleated cells stained with 4’,6-diamidino-2-phenylindole (DAPI) in blue, containing foci of $\gamma$-H2AX immunostained with Cyanine 3 (Cy3) in red.}
\label{fig:Foci_example}
\end{figure}

When a cell is exposed to radiation, a wide variety of DNA damage can occur, including double-strand breaks (DSBs), which literally means that the two strands of the double helix break in close proximity. The cell has DNA damage response mechanisms, that they are activated in the face of this type of disruption. A critical modification is that in presence of DSBs the histone variant H2AX is phosphorylated ($\gamma$-H2AX) near the break site. This modification is extended to many megabases of the chromatin allowing to be detected microscopically in form of nuclear foci by immunostaining techniques \cite{rogakou1998dna}, as seen in Figure \ref{fig:Foci_example}. Those visible fluorescent foci can then be scored manually or automatically by a computer program. Because foci yield is proportional to the absorbed radiation dose, the results from several hundred cells can be used for dose estimation analysis \cite{rothkamm2009gamma, rothkamm2013manual}.

The first part of the dose estimation method involves the calibration process. Samples of human peripheral blood lymphocytes are irradiated under controlled conditions at predetermined doses. Then the number of foci counts in peripheral blood mononucleated cells (PBMCs) is collected and the results are fitted to regression curves or surfaces, giving a calibration model. To determine the absorbed radiation dose by an exposed person, the number of $\gamma$-H2AX foci in the blood sample of that person must be evaluated. Then an inverse regression method \cite{merkle1983statistical,higueras2015}, using the calibration model and the foci counts in the patient's sample, allows to estimate the dose received by the patient.

\begin{figure}
\centering
\includegraphics[width=0.9\textwidth, height=4cm]{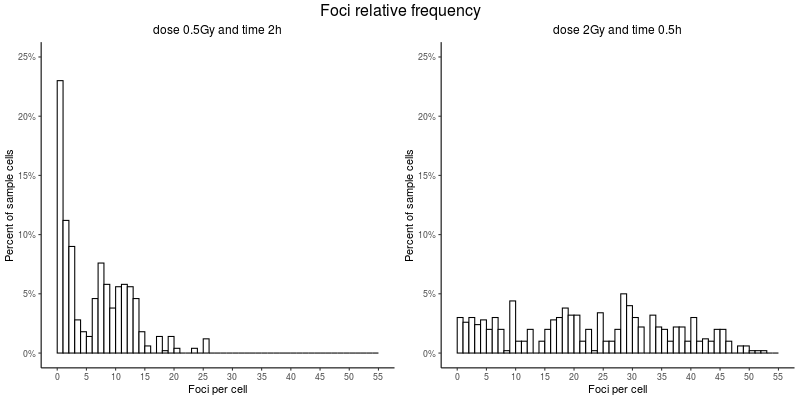}
\caption{$\gamma$-H2AX foci relative frequency found in 500 cells for dose 0.5 Gy and post-irradiation time 2h (left) and for dose 2 Gy and post-irradiation time 0.5h (right).}
\label{fig: Foci_freq}
\end{figure}

It is known that the number of DSBs depends on the amount of radiation received by cells; the more radiation absorbed, the greater the biological damage. However, the $\gamma$-H2AX foci are influenced not only by dose but also by the time from exposure \cite{moquet2017second}. As studied before, the $\gamma$-H2AX level reaches the maximum 30 minutes after exposure and then gradually decreases over hours as the cells repair the damage, returning to normal levels within 24-48 hours \cite{mariotti2013use, redon2009gamma}. For this reason, in the case of a radiological accident the utility of $\gamma$-H2AX assay is restricted to rapid usage. The distribution of the number of observed foci varies greatly depending on the time and dose, as shown in the Figure \ref{fig: Foci_freq}. As the dose is increased, fewer zeros appear because cells tend to have more foci. The number of foci, on the other hand, gets smaller over time.

Due to the great variability over time of the foci data, the calibration curves have commonly been constructed separately for different times after exposure \cite{einbeck2018statistical, chaurasia2021establishment}. From a practical point of view, this approach is somewhat unrealistic given the continuity of time and the inability to accurately determine the exact time point of irradiation. Therefore, it seems a noteworthy idea to include the time as a variable in a calibration model and to consider a three-dimensional response surface, i.e. simultaneously dose and time dependent, what was recently proposed by Lopez et al. \cite{lopez2021establishment}.

Foci $\gamma$-H2AX data are overdispersed \cite{einbeck2018statistical, rothkamm2013manual}, which means that the variance in the observed samples is higher than its mean. Due to this fact, the Poisson model, commonly used in the analysis of count data, may not be appropriate. Instead, many distributions were already proposed, in particular, negative binomial, zero-inflated Poisson and zero-inflated negative binomial models provide good results for $\gamma$-H2AX foci \cite{chilimoniuk2021countfitter}. $\gamma$-H2AX is analyzed in PBMCs, that is composed by a mixture of various types of leukocytes, but different subsets of blood cells could have different levels of $\gamma$-H2AX foci. It has been suggested that the subset of CD4+ leukocytes show 1.5 times higher level of phosphorylation than CD19+ \cite{andrievski2009response}. Although this statement requires more investigation because more evidence is necessary, it can be a plausible explanation for the variability in the distribution of foci, hence we opted to use a mixed Poisson model in this paper. Mixture models are a useful tool for modelling data that are highly diverse, and it is thought that this behaviour is due to underlying sub-populations \cite{mclachlan2019finite}. These models assume that it is not known to which subgroup a particular observation belongs, so a mixture of a few Poisson distributions with different proportions is used to describe all the data.

Moreover, many unknowns may arise in the dose estimation process, such as the impossibility of determining the exact time of irradiation or the individual response to radiation. In this scenario, the Bayesian framework may be more appropriate because it can handle many levels of uncertainty \cite{ainsbury2014review}. The Bayesian approach also allows current information to be incorporated into the inferential process, as may be the case in biodosimetry, for example when laboratories prepare a calibration curve (or surface) independent of emergency data. Therefore, some information about the distribution of the model parameters may be provided in advance of a possible accident. 

This study aims to investigate the use of biomarker $\gamma$-H2AX in the estimation of the radiation dose absorbed by an individual taking into account the uncertainty on their exposure time. In the second section, we introduce statistical strategies for dealing with the issues outlined before. In the third section, we apply these methods to $\gamma$-H2AX real data and we present the results. In the last part, we discuss some limitations of the statistical proposals and we discuss some directions for further research. 

\section{Statistical models}
\begin{figure}
	\centering
	\includegraphics[width=0.8\textwidth]{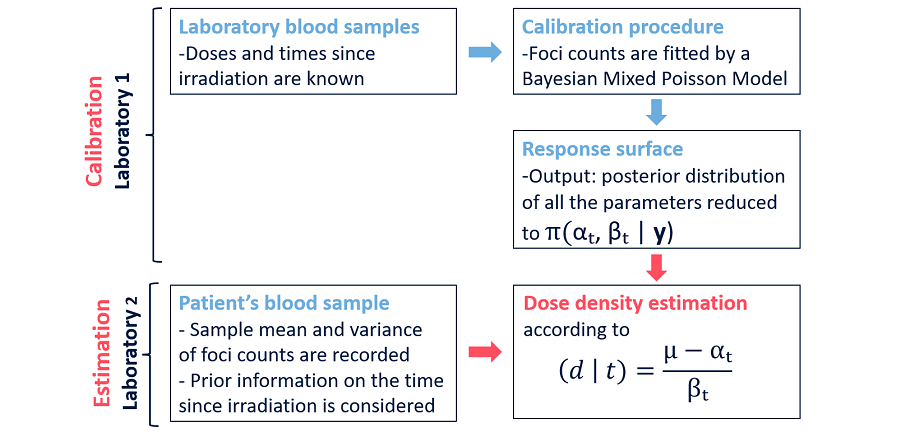}
	\caption{Scheme of the whole dose estimation procedure.}
	\label{fig: scheme}
\end{figure}

In this section, we describe the proposed statistical methodology for analysing biodosimetric foci-data. The process is divided into two parts: calibration and estimation. The calibration procedure is based on sample laboratory data, implying that the irradiation exposure occurs under monitored conditions. Calibration data were obtained irradiating $N$ peripheral mononucleated cells from one donor with radiation doses ranging from 0 to 3 Gy, and $\gamma$-H2AX foci were detected microscopically using a semi-automatic method, at different post-irradiation times from 0.5 to 24 h. The number of $\gamma$-H2AX foci, $y_i$, was recorded, obtaining a set of data $\textbf{y}=(y_1, \ldots, y_N)$, assuming conditional independence among observations. The data can be found in the supplemental material of the paper by López et al. \cite{lopez2021establishment}. These data, which serve as a type of reference for how cells behave to different doses and time points, are fitted with our statistical model to create a calibration surface. Then, blood samples from a potentially newly irradiated patient are explored, and the mean and variance of the number of foci counts in $n$ cells are recorded. In the estimation part of the process, they are used to estimate the doses supposedly received by those irradiated persons. Calibration is only required once, and the same surface response can be used to analyse many patient samples.
In fact, different laboratories can perform calibration and estimation; Figure \ref{fig: scheme} describes the process.

\subsection{Calibration}\label{calibration}
As explained in the introduction, we propose a Poisson mixture model with K components to describe the sampling distribution of the number of $\gamma$-H2AX foci per cell. We assume that observations $\textbf{y}=(y_1, \ldots, y_N)$ are conditionally independent and have been generated by the following finite mixture

\begin{equation}
(y_i \mid \boldsymbol \omega, \boldsymbol \lambda_i) \sim \sum_{k=1}^K \omega_k f(y_i \mid \lambda_{ki}),
\label{mixture}
\end{equation}

where $\boldsymbol \omega= (\omega_1, \ldots, \omega_K)$, $\boldsymbol \lambda_i=(\lambda_{1i}, \ldots, \lambda_{Ki})$ and $f(y_i \mid \lambda_{ki})= \frac{e^{-\lambda_{ki}} \lambda_{ki}^{y_i}}{y_i !}$ indicates the conditional Poisson probability of observing $y_i$ foci given $\lambda_{ki}$. The parameters $0<\omega_k<1$ represent the weight of each component of the mixture and they sum up to 1, $\sum_{k=1}^K \omega_k=1$. 

In general, the parameter $\lambda$ of a Poisson random variable can be modelled in many ways using different link functions. As mentioned earlier, time since radiation exposure has a substantial influence on the presence of the number of $\gamma$-H2AX foci. For this reason, and in line with the model proposed by \cite{lopez2021establishment}, $\lambda$ could be defined as a function of two variables, dose and time (represented by $d$ and $t$ respectively), $\lambda=\lambda(d,t)= c \cdot t^{u} + a \cdot t^{v} \cdot d,$ where $(a,c,u,v)$ are parameters. Therefore, $\lambda$ is a three-dimensional surface, although it is important to note that if time is fixed, $\lambda$ is linear with respect to $d$, which is consistent with previous findings \cite{einbeck2018statistical}.

For the mixture model with $K$ components in (\ref{mixture}), the parameters $\lambda_{ki}$ are defined separately for each Poisson component $k \in \{1,\ldots, K \}$, $\lambda_{ki}(d,t) = c_k \cdot t_i^{u_k} + a_k \cdot t_i^{v_k} \cdot d_i$ for $i={1, \ldots, N}$. Since the calibration phase uses laboratory data, for a given observed number of foci $y_i$, $t_i$ (time) and $d_i$ (dose) are known. From now on to simplify the notation, $ \boldsymbol \theta_k=(a_k,c_k,u_k,v_k)$ stands for the set of parameters for component $k$ and $\boldsymbol\theta=(\theta_1,\ldots,\theta_K)$ for all parameters in the model. In order to find the calibration surface, these parameters, together with the weights of the mixture $\boldsymbol \omega=\{\omega_1,\ldots,\omega_K\}$, should be estimated, what we propose to do within Bayesian framework.

In the Bayesian schema, all previous information about the quantities of interest are used to elicit a prior density distribution $\pi (\boldsymbol \omega, \boldsymbol \theta)$ for the parameters $(\boldsymbol \omega, \boldsymbol \theta)$. Next, the data $\textbf{y}$ are registered and the likelihood $\mathcal L(\textbf{y} \mid \boldsymbol \omega, \boldsymbol \theta)$ of $(\boldsymbol \omega, \boldsymbol \theta)$ for the data are constructed. In our case, given the conditional independence of the observations $y_i, \ldots, y_N$, the likelihood function remains,

\begin{equation*}
\mathcal L (\textbf{y} \mid \boldsymbol \omega, \boldsymbol \theta)= \prod_{i=1}^{N} \sum_{k=1}^K \omega_k f(y_i \mid c_k \cdot t_i^{u_k} + a_k \cdot t_i^{v_k} \cdot d_i). \label{eq:likel_cal}
\end{equation*}

Then the Bayes theorem provides the posterior density distribution $\pi (\boldsymbol \omega, \boldsymbol \theta \mid \textbf{y})$:

\begin{equation*}
\begin{split}
 \pi(\boldsymbol \omega, \boldsymbol \theta \mid \textbf{y}) \propto & \mathcal L(\textbf{y} \mid \boldsymbol \omega, \boldsymbol \theta) \pi(\boldsymbol \omega, \boldsymbol \theta). \label{eq:post_full}
\end{split}
\end{equation*}

As stated before, Bayesian modelling requires the specification of a prior distribution $\pi(\boldsymbol \omega, \boldsymbol \theta)$. We assume prior independence between $\boldsymbol \omega$ and $\boldsymbol \theta$ and, consequently $\pi (\boldsymbol \omega, \boldsymbol \theta)= \pi (\boldsymbol \omega) \pi (\boldsymbol \theta)$. We propose to consider non-informative uniform prior distributions for $\boldsymbol \theta$, as they will produce a minimal influence on the inference. For $\boldsymbol \omega$, we decided to use a non-informative symmetric Dirichlet distribution called Perks' prior (see \cite{berger2015overall} and supplemental material). 

The analytical derivation of the posterior distribution is intractable due to model's complexity; however, the Bayesian framework provides numerous computational methods for approximating posterior distributions. In particular, sampling algorithms, which rely on Markov chain Monte Carlo (MCMC) methods, can give a good approximation of the posterior distribution. One such method is the Gibbs sampler, in which samples from $\pi(\boldsymbol \omega, \boldsymbol \theta \mid \textbf{y})$ are constructed from the conditional posterior distribution of each element in $(\boldsymbol \omega,\boldsymbol \theta)$ given the rest of them. Posterior estimates of $(\boldsymbol \omega, \boldsymbol \theta)$ are the subsequent posterior means, approximated by the corresponding sample means of the MCMC posterior distribution.

Nonetheless, when dealing with a large number of model parameters, MCMC techniques are often extremely slow. If the absorbed radiation dose must be estimated immediately the technique of providing the results must be as quick as possible. Therefore, we propose to use the Laplace approximation method, which aims to find a Gaussian approximation to the posterior distribution, which results in a faster execution of the analysis. Assume that $\boldsymbol \phi^{\ast}=(\hat{\omega},\hat{\theta}) \in \mathbf{R}^p$ is the mode of $ \pi(\boldsymbol \omega, \boldsymbol \theta \mid \textbf{y})$, i.e.

\begin{equation*}
\phi^{\ast} = \underset{(\omega,\theta)}{\arg\max} \: \mathcal L(\textbf{y} \mid \boldsymbol \omega, \boldsymbol \theta) \pi(\boldsymbol \omega, \boldsymbol \theta).
\end{equation*}

The Laplace approximation of the posterior $ \pi(\boldsymbol \omega, \boldsymbol \theta \mid \textbf{y})$ provides a $p$-dimensional multivariate normal distribution (details can be found in supplemental material):

\begin{equation}
 \pi(\boldsymbol \omega, \boldsymbol \theta \mid \textbf{y}) \sim \mathcal{N}_p (\boldsymbol \phi^{\ast}, \hat{\Sigma}_{\hat{\omega},\hat{\theta}}),
\end{equation}

\noindent
where $\hat{\Sigma}_{\hat{\omega},\hat{\theta}}$ is the inverse of the Hessian matrix of the model evaluated at $(\hat{\omega},\hat{\theta})$ (the estimated variance-covariance matrix). The estimated parameters $(\hat{\omega},\hat{\theta})$ will be called \textit{calibration coefficients}.

For a fixed time $t$ the expected number of foci provided by the calibration surface remains,

\begin{equation}\label{mu}
\mu (d \mid t)=\sum_{k=1}^K \omega_k \lambda_{k}=\sum_{k=1}^K \omega_k (c_k \cdot t^{u_k} + a_k \cdot t^{v_k} \cdot d)=\alpha_t +\beta_t d, 
\end{equation}

where, 

\begin{equation*}
\begin{split}
\alpha_t= \sum_{k=1}^K \omega_k \cdot c_k \cdot t^{u_k}=g_1(\boldsymbol \omega, \boldsymbol \theta), \hspace{0.5cm}
\beta_t= \sum_{k=1}^K \omega_k \cdot a_k \cdot t^{v_k}=g_2(\boldsymbol \omega, \boldsymbol \theta).
\end{split}
\end{equation*}

\noindent
The derived posterior distribution of the two time-dependent parameters $(\alpha_t, \beta_t)$ is the output of the calibration process. Using the multivariate delta method, it is straightforward to see that $(\alpha_t, \beta_t)$ follows an approximated bivariate normal distribution,

\begin{equation}
\pi (\alpha_t, \beta_t \mid \textbf{y}) \dot{\sim} \mathcal{N}_2 (\textbf{g}(\hat{\omega}, \hat{\theta}), \nabla \textbf{g} \cdot \hat{\Sigma}_{\hat{\omega},\hat{\theta}} \cdot \nabla \textbf{g}^T),
\label{bivariate}
\end{equation}

where $\nabla \textbf{g}$ is the gradient of $\textbf{g}=(g_1,g_2)$ evaluated at the estimated modal point $(\hat{\omega},\hat{\theta})$.

\subsection{Estimation}
In this part, we assume that there is available a set of new observations $\textbf{x}=(x_{1},\ldots, x_{n})$, coming from a new patient, called test data. In the event of a radiation emergency, these will be the counts of foci detected in the exposed person's blood cells, for which the absorbed dose must be estimated. After a fast exploration the laboratory records the mean number of foci in the test data $\bar{x}$ and its standard deviation $s$. We can assume that the expected number $\mu$ of foci of the patient, independently of time t, follows a normal distribution $\mu\sim\mathcal{N}(\bar{x}, \frac{s^2}{n})$, where $n$ is the number of observations in the test data. This is a kind of nonparametric Bayesian estimate of $\mu$ justified in the supplemental material. 

We use the dose relation described in (\ref{mu}) to relate the patient's observed foci distribution with time $t$ and dose $d$. Then, the dose can be directly determined as

\begin{equation}
(d \mid t)=\frac{\mu-\alpha_t} {\beta_t},
\label{eq:dose}
\end{equation}

where $(d \mid t)$ means that dose is defined for a given value of $t$. Expression (\ref{eq:dose}) establishes that, when the bivariate normal approximation (\ref{bivariate}) is used for approximating the posterior distribution for $(\alpha_t, \beta_t)$, the posterior distribution of $d$ conditioned to $t$, $\pi(d \mid t,\textbf{y}, \textbf{x})$ can be approximated by the ratio of two dependent normal variables (see \cite{pham2006density} and supplemental material). Given a prior information on the post-irradiation time, summarized with the prior distribution $\pi(t)$, the joint posterior density of the time and dose of the new irradiated person can be expresses as $\pi(d,t \mid \textbf{y}, \textbf{x}) =\pi(d \mid t, \textbf{y}, \textbf{x})\pi(t)$. 

The marginal posterior density of the dose $\pi(d \mid \textbf{y}, \textbf{x})$ can be done by simulation accordingly with the following algorithm:

\begin{enumerate}
	\item Generate a value $t^*$ of the prior distribution $\pi(t)$.
	\item Generate a value of $(\alpha_{t^*},\beta_{t^*})$ from $\pi(\alpha_t, \beta_t \mid \textbf{y})$. It can be done using Gibbs sampler or using the Laplace approximation. In the last case, $(\alpha_{t^*},\beta_{t^*})$ is generated using the bivariate normal distribution (\ref{bivariate}).
	\item Generate a value $\mu^*$ of the normal distribution $\mathcal{N}(\bar{x}, \frac{s^2}{n})$. Use expression (\ref{eq:dose}) with the inputs $\mu^*$, $\alpha_{t^*}$ and $\beta_{t^*}$ for obtaining an estimated value of the dose $d^*$ and record it.
	\item Repeat steps 1 to 4 many times (at least 10,000). 
\end{enumerate}

The simulated values $d^*$ from $\pi(d \mid \textbf{y}, \textbf{x})$ allow to obtain the median and credibility intervals of the dose received by the patient. It is worth to mention that when the Laplace approximation is used the density $\pi(d \mid \textbf{y}, \textbf{x})$ can also be obtained by numerical integration of $\pi(d,t\mid\textbf{y}, \textbf{x})$ with respect to $t$, see supplemental material for details. 

\section{Materials and results} \label{sec:results}
Except for the three data sets chosen to test the models' performance, we decided to use all of the data published by \cite{lopez2021establishment} as calibration data.
Radiation doses range from 0 to 3 Gy, with post-irradiation times ranging from 0.5 to 24 hours.
The supplemental material contains all of the data used in this study as well as the R scripts used to obtain the results.
Automatic microscopy was used to count the number of foci in 500 cells for each dose and time.
The foci frequencies increase with increasing doses, as expected, but they decrease throughout time, reaching their lowest level after 24 hours.

The calibration data were fitted to mixture Poisson models with $K=2,3,4$ and 5 components. Based on the Akaike information criterion (AIC \cite{akaike1998information}), we found that the best model was the mixture Poisson model with $K=4$ components. Moreover, we chose a model where the parameter $u_{k}$ is equal in all four components and from now on it will be denoted as $u$. It means that for further analysis the mixture Poisson model with 4 components was chosen with mean given by 
 $$\lambda_k(d,t)= c_k \cdot t^{u} + a_k \cdot t^{v_k} \cdot d, \hspace{1 cm} k=1,2,3,4.$$

leading to a 16 parameters model. Figure S1 in supplemental material shows the profile of the calibration surface. 

We used the test data from three data sets which weren't included in the calibration data as an example of application to illustrate the methodology.
Radiation doses were 0.75, 2, and 3 Gy, respectively, with post-irradiation durations of 4, 10, and 0.5 hours. 
We assume that the laboratory that will conduct the patients' blood tests will only provide aggregate foci data, such as the sample mean and variance, rather than raw data. 

We chose to check the method's robustness by using three different prior distributions ($\pi(t)$) for each test data set. The first prior is a uniform distribution on the intervals $(0.25,0.75)$, $(3,5)$, and $(8,12)$ that is centred at the known post-irradiation times of 0.5, 4, and 10 hours, respectively. The second and third options are non-standard symmetric beta distributions (see supplemental material for details) with parameters $\alpha=5,\beta=5$ and $\alpha=100,\beta=100$, on the same intervals as the uniform priors. 

\begin{figure} 
	\centering
	\includegraphics[width=1\textwidth]{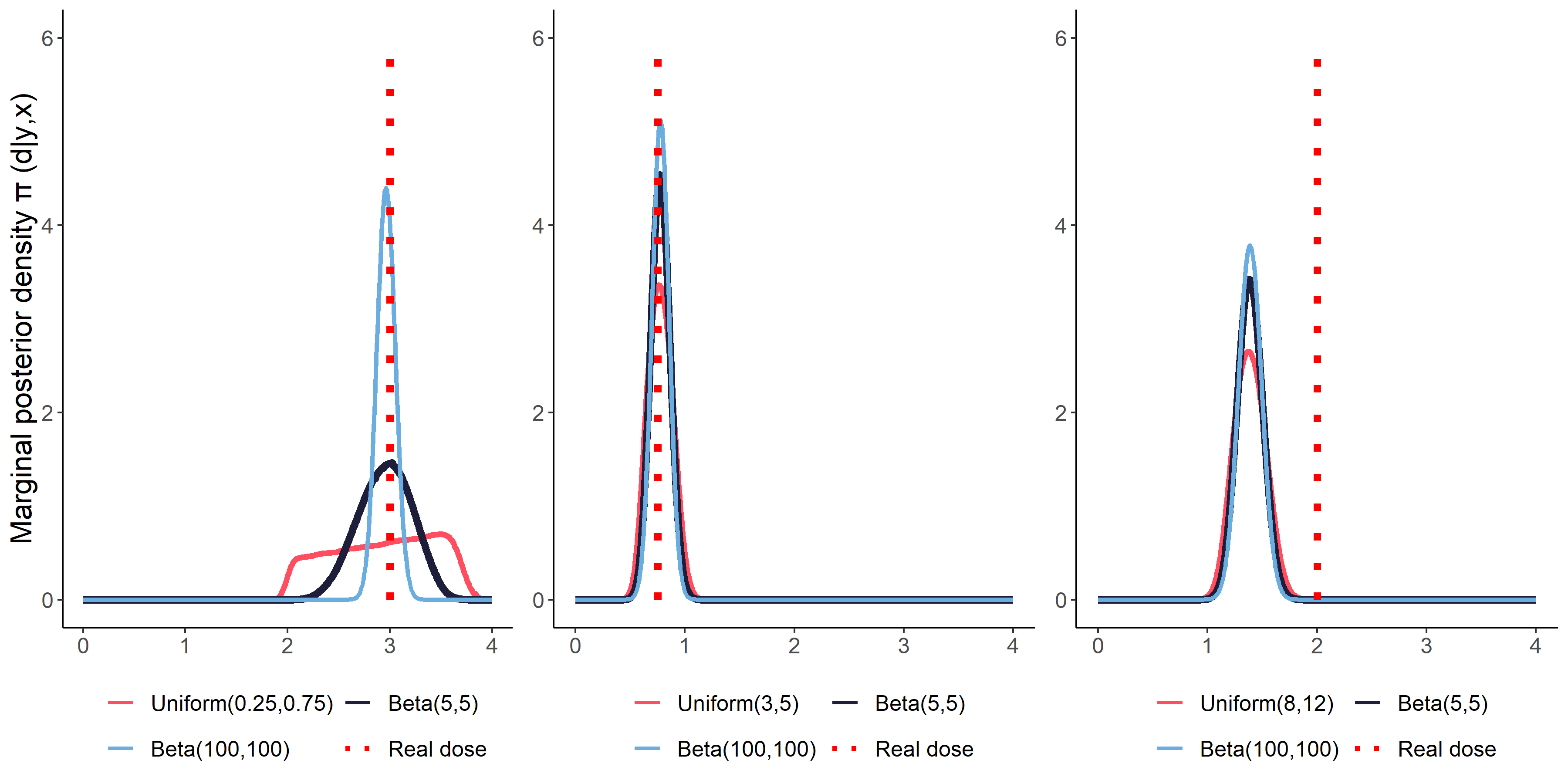}
	\caption{Marginal posterior densities $\pi(d \mid \textbf{y}, \textbf{x})$ for the test data obtained by Laplace approximation method. The prior distributions of the post-irradiation time are defined on the intervals: $(0.25,0.75),(3,5),(8,12)$ h.}
	\label{fig: posterior_density}
\end{figure}

\begin{table}[ht]
\begin{adjustwidth}{}{}
\centering 
\resizebox{\textwidth}{!}{%
\begin{tabular}{|c|c|c|c|c|c|c|c|c|}
\hline
\multicolumn{4}{|c|}{Test data} &\multicolumn{4}{c|}{Dose estimation} \\
\hline
Donor & Dose Gy & Time h & Foci Mean $\pm$ SE & \multicolumn{4}{c|}{ Prior distribution of time } \\
\hline
\multicolumn{4}{|c|}{} & Time interval h & Uniform & Beta(5,5) & Beta(100,100) \\ 
 \hline
1 & 3.00 & 0.5 & 28.612 $\pm$ 0.525 & (0.25,0.75) &\makecell{2.93\\ (2.044,3.704)} & \makecell{2.955\\ (2.43,3.434)} & \makecell{2.966\\ (2.785,3.139)} \\ 
  \hline
2 & 0.75 & 4.0 & 4.072 $\pm$ 0.230 & (3,5) &\makecell{0.771\\ (0.567,0.989)} & \makecell{0.773\\ (0.605,0.948)} & \makecell{0.774\\ (0.622,0.927)} \\ 
  \hline
1 & 2.00 & 10.0 & 4.036 $\pm$ 0.198 & (8,12) &\makecell{1.38\\ (1.115,1.662)} & \makecell{1.382\\ (1.159,1.614)} & \makecell{1.383\\ (1.178,1.591)} \\ 
  \hline
\end{tabular}}
\end{adjustwidth}
\caption{Test data and dose estimates (mean and credible interval) obtained by Laplace approximation method.}
\label{results:Laplace}
\end{table}

To estimate the posterior marginal distribution for the dose $\pi(d \mid \textbf{y}, \textbf{x})$ we have applied the Laplace approximation and marginalising $\pi(d,t \mid \textbf{y}, \textbf{x})$. The analysis was performed using R and JAGS program and the codes can be found in supplemental materials. The results of the estimation are presented in Table \ref{results:Laplace} and the posterior densities $\pi(d \mid \textbf{y}, \textbf{x})$ can be seen in Figure \ref{fig: posterior_density}. The goodness of the Laplace approximation with respect to the Gibbs sampler has been checked providing the same results up to two decimal places (see Table S1 in supplemental material).

It can be seen that the model performs well for the two first test data, where the estimated dose is very close to the real one independently of the prior distribution of the time since irradiation considered. However, for the third data set the dose is underestimated. Not surprisingly, the credible intervals are the narrowest for the prior beta distribution with $\alpha=100,\beta=100$.

\section{Discussion}
The methods of biodosimetry are very useful in both small and large-scale accidents \cite{world2011cytogenetic}. However, in both cases it is important to timely determine the radiation dose to those exposed, as it hopefully could lead to a prompt and effective treatment. Therefore the method of dose estimation should be quick and easy to use by people who provide help during emergencies \cite{vinnikov2010limitations}.

The main advantage of $\gamma$-H2AX assay, comparing to others biomarkers, is its speed, because it does not require a long process of culturing cells and can provide results within a few hours of receiving a blood sample \cite{ainsbury2014inter}, what makes $\gamma$-H2AX biomarker a good tool for rapid triage in case of a mass casualty event \cite{moquet2017second}. In addition to the duration of the biological part of the procedure, we suggest also taking into account the speed of generating estimation results, because the more complex the model, the longer it takes to execute it.

The method proposed by \cite{lopez2021establishment} does not use the raw data for constructing the calibration surface and does not take into account previous knowledge about the post-irradiation time. The Bayesian method proposed in this study, which is based on the Laplace approximation and numerical integration, is quick and accurate, and requires no simulations. When compared to the Gibbs sampling method, it produces nearly identical results, but the time required to estimate them is significantly longer. About 5,000 iterations of the Gibbs sampling method were done, and the results took more than 39 hours with these settings. Applying Laplace approximation allows for a significant reduction in the computation time needed to estimate the dose as the results of this method are immediate. 

The outputs of the calibration procedure are the estimated values for 16 parameters, the calibration coefficients $(\hat{\omega},\hat{\theta})$, and the 136 estimated values of the variance-covariance matrix $\hat{\Sigma}_{\hat{\omega},\hat{\theta}}$. By using our methodology and the 152 figures generated by Laboratory 1, other labs (or the same Laboratory 1) can analyse a patient's blood sample and determine the dose of radiation they received.

\section*{Funding}
This work was supported by the Consejería de Educación, Cultura y Deportes (Junta de Comunidades de Castilla-La Mancha (Spain)) [the Project MECESBAYES (SBPLY/17/180501/000491)]; Ministerio de Ciencia e Innovación (Spain) [research grants PID2019-106341GB-I00, RTI2018-096072-B-I00]; the Spanish Consejo de Seguridad Nuclear [BOE-A-2019-311]; and the Spanish State Research Agency [the Severo Ochoa and Marıa de Maeztu Program for Centers and units of Excellence in R\&D (CEX2020-001084-M)].

\noindent
\Large Supplemental material for the article \textbf{Improving radiation dose estimation using the $\gamma$-H2AX biomarker}
\normalsize
\\
\\
Dorota Młynarczyk\textsuperscript{1,*}, Pedro Puig\textsuperscript{1,2}, Carmen Armero\textsuperscript{3}, Virgilio Gómez-Rubio\textsuperscript{4}, Joan F. Barquinero\textsuperscript{5}, Mònica Pujol-Canadell\textsuperscript{5}
\vspace{0.5cm}\\
\small
\textsuperscript{1} Departament de Matemàtiques, Universitat Autònoma de Barcelona, Bellaterra, 08193, Barcelona. \\
\textsuperscript{2} Centre de Recerca Matemàtica, Bellaterra, 08193 Barcelona. \\
\textsuperscript{3} Departament d’Estadística i Investigació Operativa, Universitat de València, 46100 València. \\
\textsuperscript{4} Department of Mathematics, School of Industrial Engineering, Universidad de Castilla-La Mancha, 02071 Albacete. \\
\textsuperscript{5} Unitat d’Antropologia Biològica, Departament de Biologia Animal, Biologia Vegetal i Ecologia, Universitat Autònoma de Barcelona, Bellaterra, 08193 Barcelona. \\
\textsuperscript{*} Correspondence to: dorotaanna.mlynarczyk@uab.cat

\subsection*{Perk's distribution}
The Dirichlet distribution of order $K \geq 2$ is parameterised by a vector of positive real numbers $\mathbf{\alpha}=(\alpha_1, \alpha_2, \ldots, \alpha_M)$ called concentration parameters. Its probability density function is given by
\[f( \omega_1,\ldots, \omega_K \mid \mathbf{\alpha} )= \frac{\Gamma (\sum_{i=1}^{K} \alpha_i)}{ \prod_{i=1}^{K} \Gamma (\alpha_i)} \prod_{i=1}^{K}\omega_{i}^{\alpha_i-1},\]
where $\omega_i \geq 0$ and $\sum_{i=1}^K \omega_i=1.$ As a result of Alvares et al. (2018), we chose the Perks' prior. This is a Dirichlet distribution where all of the parameters are equal to $1/K$. This prior was proposed by Perks (1947), but Berger et al. (2015) obtained it as the reference distance prior. 

\subsection*{Laplace approximation}
Assume that $\boldsymbol \phi^{\ast}=(\hat{\omega},\hat{\theta}) \in \mathbf{R}^p$ is the mode of the posterior density $ \pi(\boldsymbol \omega, \boldsymbol \theta \mid \textbf{y})$. Taking the second order Taylor expansion of $f(\boldsymbol \phi)=f(\boldsymbol \omega, \boldsymbol \theta)=\log ( \pi(\boldsymbol \omega, \boldsymbol \theta \mid \textbf{y}))$ centered on $\boldsymbol \phi^{\ast}$, we get:

$$
f(\boldsymbol \phi) \approx f( \boldsymbol \phi^{\ast})-\frac{1}{2}(\boldsymbol \phi-\boldsymbol \phi^{\ast})^T \boldsymbol H^{\ast}( \boldsymbol \phi- \boldsymbol \phi^{\ast}), 
$$
\noindent
where $\boldsymbol H^{\ast}$ is the Hessian matrix, the matrix of second-order partial derivatives of $f(\boldsymbol \phi)$, evaluated at $\phi^{\ast}$. Taking the exponential it gives, 

$$
 \pi(\boldsymbol \omega, \boldsymbol \theta \mid \textbf{y}) \propto \mathcal L(\textbf{y} \mid \boldsymbol \phi^{\ast}) \pi(\boldsymbol \phi^{\ast}) \exp [-\frac{1}{2}(\boldsymbol \phi-\boldsymbol \phi^{\ast})^T \boldsymbol H^{\ast}( \boldsymbol \phi- \boldsymbol \phi^{\ast})]. 
$$

\noindent
The right side of the equation is a $p$-dimensional multivariate normal distribution, so the Laplace approximation provides a Gaussian approximation of the posterior, that is, 
$$
 \pi(\boldsymbol \omega, \boldsymbol \theta \mid \textbf{y}) \sim \mathcal{N} (\boldsymbol \phi^{\ast}, \boldsymbol H^{\ast-1}).
$$
\noindent
Thus it remains to determine the mode $\boldsymbol \phi^{\ast}$ of the posterior density $ \pi(\boldsymbol \omega, \boldsymbol \theta \mid \textbf{y})$. This procedure is known as maximum a posteriori (MAP) estimation and it is defined as,
$$
 \phi^{\ast} = \underset{(\omega,\theta)}{\arg\max} \: \mathcal L(\textbf{y} \mid \boldsymbol \omega, \boldsymbol \theta) \pi(\boldsymbol \omega, \boldsymbol \theta).
$$
\noindent
Note that if flat distributions are chosen as prior distributions of model parameters, $\pi(\boldsymbol \omega, \boldsymbol \theta)\sim 1$, this estimator $\phi^{\ast}$ coincides with the maximum likelihood estimator (MLE), i.e. the model parameters which maximize the likelihood function. 

\subsection*{A nonparametric Bayesian estimate of $\mu$}

Let $\textbf{x}$ a vector of $n$ independent observations coming from a distribution parametrized by its population mean $\mu$, i.e. $f(x|\mu)$, with likelihood function,
$$ \mathcal L(\textbf{x}|\mu)=\prod_{i=1}^n f(x_i|\mu).$$  
It is worth to mention that $\mu$ can also depend of other parameters. The posterior distribution of $\mu$, given the prior $\pi(\mu)$, remains 

$$
\pi(\mu \mid \textbf{x}) \propto \mathcal L(\textbf{x} \mid \mu) \pi(\mu)\,.$$
It is known that for large $n$ and under commonly satisfied regularity assumptions, the posterior density can be approximated as,

\begin{equation}
\pi(\mu \mid \textbf{x}) \approx \mathcal{N}(\hat\mu, [I(\hat{\mu})]^{-1}),
\label{app1}
\end{equation}
where $\hat\mu$ is the maximum likelihood estimator (MLE) of $\mu$ and $I(\hat{\mu})$ is the Fisher information quantity evaluated at $\hat\mu$. It is remarcable that this asymptotic result is independent of the chosen prior $\pi(\mu)$ (see Berger et al. (2015), Result 8, p. 224). 

It is said that distributions satisfy the Gauss' principle if the MLE of $\mu$ is the sample mean, i.e. $\hat\mu=\bar x$ (see Puig (2008) and the references therein). For instance, the Gauss' principle is satisfied by the Normal, Poisson, and Negative Binomial distributions. In particular, it can be demonstrated that the finite mixtures of Poisson distributions satisfy the Gauss' principle using the findings of Bondesson (1997).  For these distributions expression (\ref{app1}) remains,     
$$\pi(\mu \mid \textbf{x}) \approx \mathcal{N}(\bar{x}, \hat\sigma^2/n),$$
where $\hat\sigma^2$ is the MLE of the variance $Var(x_i)=\sigma^2$. Finally, we replace $\hat\sigma^2$ by the sample variance that is a robust estimator of $\sigma^2$, obtaining,
$$\pi(\mu \mid \textbf{x}) \approx \mathcal{N}(\bar{x}, s^2/n).$$  

Due to technical limitations or a lack of a sufficient sample size, Laboratory 2 could be unable to fit a K-mixture of Poisson distributions. However, our Bayesian approach for calculating the received dose of radiation can be used by simply recording the mean number of foci found in test data, $\bar{x}$, and its standard deviation $s$ (or variance $s^2$).        

\subsection*{Ratio of two dependent normal variables}

The following result can be found in Pham-Gia et al. (2006): 
\begin{theorem}
Let $(X,Y) \sim \mathcal{N}_2 (\mu_X,\mu_Y;\sigma_X,\sigma_Y;\rho)$. Then the density of $W=X / Y$ is 
\begin{equation*}
 f_W(w;\mu_X,\mu_Y;\sigma_X,\sigma_Y;\rho)=K_2 \frac{2(1-\rho^2) \sigma_X^2 \sigma_Y^2}{\sigma_Y^2 w^2-2\rho \sigma_X \sigma_Y w +\sigma_X^2} {}_1F_1(1;1/2;\theta_2(w)),
\end{equation*}
$-\infty<w<\infty$,where \\
$\theta_2(w)=\frac{[-\sigma_Y^2\mu_Xw+\rho\sigma_X\sigma_Y(\mu_Yw+\mu_X)-\mu_Y\sigma_X^2]^2}{2\sigma_X^2\sigma_Y^2(1-\rho^2)(\sigma^2_Yw^2-2\rho\sigma_X\sigma_Yw+\sigma_X^2)} \geq 0$,\\
$K_2=\frac{1}{2\pi\sigma_X \sigma_Y \sqrt{1-\rho^2}}\exp(-\frac{\sigma_Y^2\mu_X^2-2\rho\sigma_X\sigma_Y\mu_X\mu_Y+\mu_Y^2\sigma_X^2}{2(1-\rho^2)\sigma_X^2\sigma_Y^2})$ and \\
${}_1F_1(\alpha,\gamma;z)=\sum_{k=0}^{\infty}\frac{(\alpha,k)}{(\gamma,k)}\frac{z^k}{k!}$ where $(\alpha,k)=\alpha(\alpha+1)\ldots(\alpha+k-1)=\Gamma(\alpha+k)/\Gamma(\alpha)$ with $(\alpha,0)=1.$
\end{theorem}

\bigskip
\noindent In our case the dose is calculated as,
\begin{equation*}
d=\frac{\mu-\alpha_t} {\beta_t}=\frac{X}{Y},
\end{equation*}
where
$\alpha_t=\sum_{k=1}^K \omega_k \cdot c_k \cdot t^{u_k}$, $\beta_t=\sum_{k=1}^K \omega_k \cdot a_k \cdot t^{v_k}$. Moreover, $\mu$ follows a normal distribution $\mathcal{N}(\bar{\textbf{x}}, \frac{s^2}{n})$, where $\bar{\textbf{x}}$ is the mean number of foci found in test data, $s$ is the sample standard error and $n$ is the number of observations in test data. Note that $X$ and $Y$ satisfy the assumptions of the theorem where,
$\mu_X=\mu-\alpha_t$, $\mu_Y=\beta_t$, $\sigma_X=\sqrt{\frac{s^2}{n}+\mathrm{Var(\alpha_t)}}$, $\sigma_Y=\sqrt{\mathrm{Var(\beta_t)}}$, $\rho=\frac{\mathrm{Cov}(X,Y)}{\sigma_X\sigma_Y}$ and
$\mathrm{Cov}(X,Y)=\mathrm{Cov}(\mu-\alpha_t,\beta_t)=\mathrm{Cov}(\mu,\beta_t)-\mathrm{Cov}(\alpha_t,\beta_t)=-\mathrm{Cov}(\alpha_t,\beta_t)$.

The function $_1F_1(\alpha,\gamma;z)$ is the Kummer’s classical confluent hypergeometric function of first kind. For $\alpha=1$ and $\gamma=1/2$ it can be expressed as, 
\begin{align*}
{}_1F_1(1,1/2;\theta_2(w))&=\exp(\theta_2(w))\sqrt{\pi}\sqrt{\theta_2(w)}\erf(\sqrt{\theta_2(w)})+1.
\end{align*}
\subsection*{Non-standard beta distribution}
The non-standard beta distribution is a parametrized by two positive shape parameters $\alpha,\beta>0$. Its probability density function is given by 
\[f(x)= \frac{(x-p)^{\alpha-1}(q-x)^{\beta-1}}{B(\alpha,\beta)(q-p)^{\alpha+\beta-1}}, \hspace{1cm} p \leq x \leq q,  \] 
where p and q are the lower and upper bounds, respectively, of the distribution and $B(\alpha,\beta)$ is the beta function defined by the integral 
\[B(\alpha,\beta)= \int_{0}^{1}t^{\alpha-1}(1-t)^{\beta-1} \mathrm{d} t. \] 
The expectation of the non-standard beta distribution is $p+\frac{\alpha}{\alpha+\beta}(q-p)$ and the variance is $\frac{\alpha \beta}{(\alpha+\beta)^2(\alpha+\beta+1)}(q-p)^2$.

\bigskip
\subsection*{Supplemental References}

\noindent Alvares, D., Armero, C., and Forte, A. (2018). What Does Objective Mean in a Dirichlet-multinomial
Process? {\it International Statistical Review}, {\bf 86}:4, 106--118.

\medskip
\noindent Bondesson, L. (1997). A generalization of Poincare's characterization of exponential families. {\it J. Statist. Plann. Inference}, {\bf 63}:2, 147--155.

\medskip
\noindent Perks, W. (1947). Some observations on inverse probability including a new indifference rule. {\it J. Inst. Actuaries}, {\bf 73}(2), 285--334.

\medskip
\noindent Puig, P. (2008). A note on the harmonic law: A two-parameter family of distributions for ratios. {\it Statistics \& Probability Letters},
{\bf 78}(3), 320--326.

\newpage

\begin{table}[ht]
	\begin{adjustwidth}{}{}
		\centering 
		\resizebox{\textwidth}{!}{%
			\begin{tabular}{|c|c|c|c|c|c|c|c|c|}
			\hline
\multicolumn{4}{|c|}{Test data} & \multicolumn{4}{c|}{Dose estimation} \\
				\hline
				Donor & Dose Gy & Time h & Foci Mean $\pm$ SE & \multicolumn{4}{c|}{Prior distribution of time} \\
				\hline
				\multicolumn{4}{|c|}{} & Time interval h &Uniform & Beta(5,5) & Beta(100,100) \\	\hline
				\multicolumn{8}{|c|}{Laplace approximation} \\
				  \hline
1 & 3.00 & 0.5 & 28.612 $\pm$ 0.525 & (0.25,0.75) &\makecell{2.93\\ (2.044,3.704)} & \makecell{2.955\\ (2.43,3.434)} & \makecell{2.966\\ (2.785,3.139)} \\ 
   \hline
2 & 0.75 & 4.0 &  4.072 $\pm$ 0.230 & (3,5) &\makecell{0.771\\ (0.567,0.989)} & \makecell{0.773\\ (0.605,0.948)} & \makecell{0.774\\ (0.622,0.927)} \\ 
   \hline
1 & 2.00 & 10.0 &  4.036 $\pm$ 0.198 & (8,12) &\makecell{1.38\\ (1.115,1.662)} & \makecell{1.382\\ (1.159,1.614)} & \makecell{1.383\\ (1.178,1.591)} \\ 
				\hline
				\multicolumn{8}{|c|}{Gibbs sampler} \\
				\hline
				1 & 3.00 & 0.5& 28.612 $\pm$ 0.525 & (0.25,0.75) & \makecell{2.933\\ (2.05,3.705)} & \makecell{2.959\\ (2.432,3.432)} & \makecell{2.968\\ (2.801,3.137)} \\ 
				\hline
				2 & 0.75 & 4.0 & 4.072 $\pm$ 0.230 & (3,5) & \makecell{0.774\\ (0.574,0.984)} & \makecell{0.773\\ (0.608,0.95)} & \makecell{0.773\\ (0.621,0.919)} \\ 
				\hline
				1 & 2.00 & 10.0 & 4.036 $\pm$ 0.198 & (8,12) & \makecell{1.381\\ (1.124,1.656)} & \makecell{1.379\\ (1.158,1.602)} & \makecell{1.384\\ (1.191,1.584)} \\ 
				\hline
		\end{tabular}}
	\end{adjustwidth}
	\caption{Test data and dose estimates (mean and credible interval) obtained by Laplace approximation method and Gibbs sampler.}
	\label{table: results_all}
\end{table}

\begin{figure*} 
\centering
\includegraphics[width=0.8\textwidth]{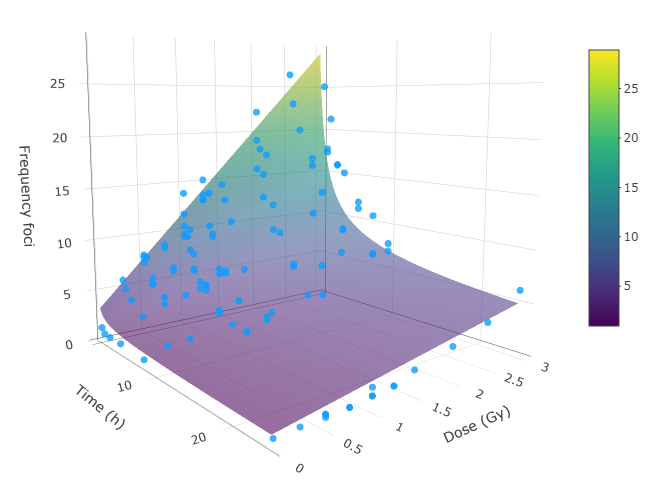}
\caption{Surface calibration model, calibration data points marked in blue.}
\label{fig: surface}
\end{figure*}

\begin{thebibliography}{10}

\bibitem{rothkamm2009gamma}
Kai Rothkamm and Simon Horn.
\newblock gamma-h2ax as protein biomarker for radiation exposure.
\newblock {\em Ann Ist Super Sanita}, 45(3):265--71, 2009.

\bibitem{rogakou1998dna}
Emmy~P Rogakou, Duane~R Pilch, Ann~H Orr, Vessela~S Ivanova, and William~M
  Bonner.
\newblock Dna double-stranded breaks induce histone h2ax phosphorylation on
  serine 139.
\newblock {\em Journal of biological chemistry}, 273(10):5858--5868, 1998.

\bibitem{rothkamm2013manual}
Kai Rothkamm, Stephen Barnard, Elizabeth~A Ainsbury, Jenna Al-Hafidh,
  Joan-Francesc Barquinero, Carita Lindholm, Jayne Moquet, Marjo
  Per{\"a}l{\"a}, Sandrine Roch-Lef{\`e}vre, Harry Scherthan, et~al.
\newblock Manual versus automated $\gamma$-h2ax foci analysis across five
  european laboratories: Can this assay be used for rapid biodosimetry in a
  large scale radiation accident?
\newblock {\em Mutation Research/Genetic Toxicology and Environmental
  Mutagenesis}, 756(1-2):170--173, 2013.

\bibitem{merkle1983statistical}
W~Merkle.
\newblock Statistical methods in regression and calibration analysis of
  chromosome aberration data.
\newblock {\em Radiation and environmental biophysics}, 21(3):217--233, 1983.

\bibitem{higueras2015}
Manuel Higueras, Pedro Puig, Elizabeth~A Ainsbury, and Kai Rothkamm.
\newblock A new inverse regression model applied to radiation biodosimetry.
\newblock {\em Proc. R. Soc. A.}, 471(20140588), 2015.

\bibitem{moquet2017second}
Jayne Moquet, Stephen Barnard, Albena Staynova, Carita Lindholm, Oct{\'a}via
  Monteiro~Gil, Vanda Martins, Ute R{\"o}{\ss}ler, Anne Vral, Charlot
  Vandevoorde, Maria Wojew{\'o}dzka, et~al.
\newblock The second gamma-h2ax assay inter-comparison exercise carried out in
  the framework of the european biodosimetry network (reneb).
\newblock {\em International journal of radiation biology}, 93(1):58--64, 2017.

\bibitem{mariotti2013use}
Luca~G Mariotti, Giacomo Pirovano, Kienan~I Savage, Mihaela Ghita, Andrea
  Ottolenghi, Kevin~M Prise, and Giuseppe Schettino.
\newblock Use of the $\gamma$-h2ax assay to investigate dna repair dynamics
  following multiple radiation exposures.
\newblock {\em PloS one}, 8(11):e79541, 2013.

\bibitem{redon2009gamma}
Christophe~E Redon, Jennifer~S Dickey, William~M Bonner, and Olga~A
  Sedelnikova.
\newblock $\gamma$-h2ax as a biomarker of dna damage induced by ionizing
  radiation in human peripheral blood lymphocytes and artificial skin.
\newblock {\em Advances in Space Research}, 43(8):1171--1178, 2009.

\bibitem{einbeck2018statistical}
Jochen Einbeck, Elizabeth~A Ainsbury, Rachel Sales, Stephen Barnard, Felix
  Kaestle, and Manuel Higueras.
\newblock A statistical framework for radiation dose estimation with
  uncertainty quantification from the $\gamma$-h2ax assay.
\newblock {\em PloS one}, 13(11):e0207464, 2018.

\bibitem{chaurasia2021establishment}
Rajesh~Kumar Chaurasia, NN~Bhat, Neeraj Gaur, KB~Shirsath, UN~Desai, and
  BK~Sapra.
\newblock Establishment and multiparametric-cytogenetic validation of
  60co-gamma-ray induced, phospho-gamma-h2ax calibration curve for rapid
  biodosimetry and triage management during radiological emergencies.
\newblock {\em Mutation Research/Genetic Toxicology and Environmental
  Mutagenesis}, 866:503354, 2021.

\bibitem{lopez2021establishment}
Juan~S L{\'o}pez, M{\`o}nica Pujol-Canadell, Pedro Puig, Montserrat Ribas,
  Pablo Carrasco, Gemma Armengol, and Joan~F Barquinero.
\newblock Establishment and validation of surface model for biodosimetry based
  on $\gamma$-h2ax foci detection.
\newblock {\em International Journal of Radiation Biology}, 98:1:1--10, 2022.

\bibitem{chilimoniuk2021countfitter}
Jaros{\l}aw Chilimoniuk, Alicja Gosiewska, Jadwiga S{\l}owik, Romano Weiss,
  P~Markus Deckert, Stefan R{\"o}diger, and Micha{\l} Burdukiewicz.
\newblock countfitter: efficient selection of count distributions to assess dna
  damage.
\newblock {\em Annals of Translational Medicine}, 9(7), 2021.

\bibitem{andrievski2009response}
Andrei Andrievski and Ruth~C Wilkins.
\newblock The response of gamma-h2ax in human lymphocytes and lymphocytes
  subsets measured in whole blood cultures.
\newblock {\em International journal of radiation biology}, 85(4):369--376,
  2009.

\bibitem{mclachlan2019finite}
Geoffrey~J McLachlan, Sharon~X Lee, and Suren~I Rathnayake.
\newblock Finite mixture models.
\newblock {\em Annual review of statistics and its application}, 6:355--378,
  2019.

\bibitem{ainsbury2014review}
Elizabeth~A Ainsbury, Volodymyr~A Vinnikov, Pedro Puig, Manuel Higueras,
  Nataliya~A Maznyk, David~C Lloyd, and Kai Rothkamm.
\newblock Review of bayesian statistical analysis methods for cytogenetic
  radiation biodosimetry, with a practical example.
\newblock {\em Radiation protection dosimetry}, 162(3):185--196, 2014.

\bibitem{berger2015overall}
James~O Berger, Jose~M Bernardo, and Dongchu Sun.
\newblock Overall objective priors.
\newblock {\em Bayesian Analysis}, 10(1):189--221, 2015.

\bibitem{pham2006density}
Thu Pham-Gia, Noyan Turkkan, and E~Marchand.
\newblock Density of the ratio of two normal random variables and applications.
\newblock {\em Communications in Statistics-Theory and Methods},
  35(9):1569--1591, 2006.

\bibitem{akaike1998information}
Hirotogu Akaike.
\newblock Information theory and an extension of the maximum likelihood
  principle.
\newblock In {\em Selected papers of hirotugu akaike}, pages 199--213.
  Springer, 1998.

\bibitem{world2011cytogenetic}
World~Health Organization et~al.
\newblock Cytogenetic dosimetry: applications in preparedness for and response
  to radiation emergencies.
\newblock Technical report, International Atomic Energy Agency, 2011.

\bibitem{vinnikov2010limitations}
Volodymyr~A Vinnikov, Elizabeth~A Ainsbury, Nataliya~A Maznyk, David~C Lloyd,
  and Kai Rothkamm.
\newblock Limitations associated with analysis of cytogenetic data for
  biological dosimetry.
\newblock {\em Radiation research}, 174(4):403--414, 2010.

\bibitem{ainsbury2014inter}
Elizabeth~A Ainsbury, Jenna Al-Hafidh, Ainars Bajinskis, Stephen Barnard,
  Joan~Francesc Barquinero, Christina Beinke, Virginie De~Gelder, Eric
  Gregoire, Alicja Jaworska, Carita Lindholm, et~al.
\newblock Inter-and intra-laboratory comparison of a multibiodosimetric
  approach to triage in a simulated, large scale radiation emergency.
\newblock {\em International journal of radiation biology}, 90(2):193--202,
  2014.

\end{thebibliography}
\end{document}